\documentclass{article}
\usepackage{graphicx}
\usepackage{bm}
\usepackage{amsmath}

\begin{document}

\title{\textbf{FLUCTUATIONS\ \,RELATIONS\ \,for\ \,SEMICLASSICAL\ \,SINGLE-MODE\
\,LASER }}
\author{Rapha$\mathrm{\ddot{e}}$l Chetrite \\
\\
Laboratoire de Physique, C.N.R.S., ENS-Lyon, Universit\'e de Lyon, \\
46 All\'ee d'Italie, 69364 Lyon, France}
\date{}
\maketitle

\abstract{\noindent Over last decades, the study of laser fluctuations
has shown that laser theory may be regarded as a prototypical
example of a nonlinear nonequilibrium problem. The present paper discusses
the fluctuation relations, recently derived in nonequilibrium statistical
mechanics, in the context of the semiclassical laser theory.}

\section{Introduction}

Nonequilibrium statistical mechanics aims at a statistical description of
closed and open systems evolving under the action of time-dependent
conservative forces or under time-independent or time dependent
non-conservative ones. \textbf{Fluctuation relations} are robust identities
involving the statistics of entropy production or performed work in such
systems. They hold arbitrarily far from thermal equilibrium, reducing close
to equilibrium to Green-Kubo or fluctuation-dissipation relations usually
obtained in the scope of linear response theory \cite
{Han,Risk,Zwan,HaySas,Fal,Che1}. In a previous paper \cite{Che}, we
presented a unified approach to fluctuation relations in classical
nonequilibrium systems described by diffusion processes. We traced the
origin of different fluctuation relations to the freedom of choice of the
time inversion. The purpose of this paper is to illustrate the results of 
\cite{Che} on the example of a phenomenological model of laser described by
a stochastic differential equation. The semiclassical theory of laser
describes the regime where, due to a large number of photons in the laser
cavity, one may treat the electrical field classically, but the two level
atoms are treated quantum mechanically \cite{Sargent,Haken}. The dynamical
behavior of a single mode laser is then described by the equation of motion
for the complex amplitude of the electric field $E_t$: 
\begin{equation}
\frac{dE}{dt}=(a_{t}-bE\bar{E})E,  \label{1}
\end{equation}
where $\bar{E}_t$ is the complex conjugate of $E_t$. The function $a_{t}$ is
called the net gain coefficient and it takes into account the coherent
emission and absorption of atoms and the losses. In the general case, $a_{t}$
may have an explicit dependence on time. $b$ is called the self-saturation
coefficient. In most instances, it has a positive real part. There exist
cases (with absorber) \cite{Sargent75} where $b$ has a negative real part,
but we shall not consider them below. If the resonance frequency $\omega _{c}
$ of the laser cavity and the atomic frequency $\omega _{a}$ are exactly
tuned then both $a_{t}$ and $b$ are real. In the case of detuning \cite
{Risken74}, $a_{t}$ and $b$ are both complex. The equation of motion (\ref{1}%
) describes the dynamical behavior of the laser field in a completely
deterministic manner with the properties like coherence or spectral width
lying outside the domain of the theory. The key to the understanding of such
questions resides in the fluctuations of the electric field which are caused
by random spontaneous atomic emissions. Such fluctuations may be accounted
for by replacing Eq.\thinspace (\ref{1}) by the stochastic differential
equation 
\begin{equation}
\frac{dE}{dt}=(a_{t}-bE\bar{E})E+\eta (t,E,\bar{E}),
\end{equation}
with the noise\ $\eta (t,E,\bar{E})$ mimicking the effect of the random
spontaneous emission of atoms in other modes, a purely quantum effect
neglected in the semiclassical theory, but also the effect of vibrations of
the cavity\cite{Sargent,Haken}. We shall take $\eta (t,E,\bar{E})$ as a
random Gaussian field with zero mean and delta-correlated in time. In the
following, we shall look at two possible forms for $\eta $, one additive and
the other one multiplicative. The present paper is organised as follows. In
Sect.\thinspace 2, we recall the main results of \cite{Che}. In
Sect.\thinspace 3.1, we study the most elementary model of laser: the
stationnary tuned laser with an additive noise, and show that its dynamics
satisfies the detailed balance. Sect.\thinspace 3.2 is devoted to the
fluctuation relations for a non-stationnary tuned laser. In Sect.\thinspace
4.1, we examine the case of a stationnary laser with detuning. The detailed
balance is broken here, but we show that its slight generalisation, the
modified detailed balance, still holds. In Sect 4.2, we study the
non-stationnary detuned case. In Sects.\thinspace 5, we look at a slightly
different case with the multiplicative noise. \ \vskip0.3cm

\noindent {\textbf{Acknowledgements}}. \ The author thanks Fran\c{c}ois
Delduc and Krzysztof Gawedzki for encouragement, and Patrick Loiseau for his
help in the numerical computation of Sect.\,5.2. \vskip0.1cm

\section{Fluctuation relation in diffusive systems \protect\cite{Che}}

In \cite{Che}, we dealt with arbitrary diffusion\ processes in $%
\boldsymbol{R^{d}}$ defined by stochastic differential equation (SDE) 
\begin{equation}
\dot{x}\ =\ u_{t}(x)+v_{t}(x),  \label{SDE}
\end{equation}
where $\dot{x}\ \equiv \frac{dx}{dt}$ and, on the right hand side, $u_{t}(x) 
$ is a time-dependent deterministic vector field (a drift), and $v_{t}(x)$
is a Gaussian random vector field with mean zero and covariance : 
\begin{equation}
\left\langle v_{t}^{i}(x)v_{s}^{j}(y)\right\rangle =\delta
(t-s)D_{t}^{ij}(x,y).  \label{covD}
\end{equation}
For the process solving the SDE (\ref{SDE}) defined using the Stratonovich
convention, we showed a detailed fluctuation relation (DFR): : 
\begin{equation}
\mu _{0}(dx)\,P_{0,T}(x;dy,dW)\,\exp (-W)=\mu _{0}^{\prime }(dy^{\ast
})\,P_{0,T}^{^{\prime }}(y^{\ast };dx^{\ast },d(-W)),  \label{DFR}
\end{equation}
where:

\begin{itemize}
\item  $\mu _{0}(dx)=\exp (-\varphi _{0}(x))dx$ is the initial distribution
of the original (forward) process,

\item  $\mu _{0}^{\prime }(dx)=\exp (-\varphi _{0}^{\prime }(x))dx$ is the
initial distribution of the backward process obtained from the forward
process by applying a time inversion (see below),

\item  $P_{0,T}(x;dy,dW)$ is the joint probability distribution of the time $%
T$ position $x_{\hspace{-0.02cm}_{T}}$ of the forward process starting at
time zero at $x$ and of the functional $W_{\hspace{-0.03cm}_{T}}[x]$ of the
process (to be given later) that has the interpretation of the entropy
production.

\item  $P_{0,T}^{^{\prime }}(x;dy,dW)$ is the similar joint probability
distribution for the backward process.
\end{itemize}

\noindent The key behind the DFR (\ref{DFR}) is the action of the time
inversion on the forward system. First, the time inversion acts on time and
space by an involutive transformation $\,(t,x)\rightarrow (t^{\ast
}=T-t,x^{\ast })$. Second, to recover a variety of fluctuation relations
discussed in the literature \cite
{Kurchan,LebowSp,Crooks1,Crooks2,Jarz4,SpS,CHCHJAR}, we allow for a
non-trivial behaviour of the drift $u_{t}$ under the time-inversion dividing
it into two parts: 
\begin{equation}
u_{t}=u_{t,+}+u_{t,-}
\end{equation}
with $u_{t,+}$ transforming as a vector field under time inversion, i.e. $%
u_{t^{\ast },+}^{\prime i}(x^{\ast })=+(\partial _{k}x^{\ast
,i})(x)\,u_{t,+}^{k}(x)$, and $u_{t,-}$ transforming as a pseudo-vector
field, i.e. $u_{t^{\ast },-}^{\prime i}(x^{\ast })=-(\partial _{k}x^{\ast
,i})(x)\,u_{t,-}^{k}(x)$. The random field $v_{t}$ may be transformed with
either rule: $v_{t^{\ast }}^{\prime i}(x^{\ast })=\pm (\partial _{k}x^{\ast
,i})(x)v_{t}^{k}(x)$. By definition, the backward process satisfies then the
SDE 
\begin{equation}
\dot{x}\ =\ u_{t}^{\prime }(x)+v_{t}^{\prime }(x)
\end{equation}
taken again with the Stratonovich convention. The functionnal $W_{\hspace{%
-0.02cm}_T}$ which appears in the DFR depends explicitely on the functions $%
\varphi _{0}$, $\varphi _{0}^{\prime }$ and on the time inversion and has
the explicit form: 
\begin{equation}
W_{\hspace{-0.02cm}_T}=\Delta_{\hspace{-0.02cm}_T}\varphi +\int_{0}^{T}%
\hspace{-0.1cm}J_{t}\,dt,  \label{WT}
\end{equation}
where $\Delta_{\hspace{-0.02cm}_T}\varphi =\varphi _{T}(x_{T})-\varphi
_{0}(x_{0})$ with 
\begin{equation}
\mu _{\hspace{-0.01cm}_T}(dx)\equiv \exp (-\varphi _{\hspace{-0.02cm}_T}(x))
dx\equiv \exp (-\varphi_{0}^{\prime }(x^{\ast }))dx^{\ast }=\mu _{0}^{\prime
}(dx^{\ast }),  \label{defu}
\end{equation}
and where 
\begin{equation}
J_{t}=2\widehat{u}_{t,+}\cdot d_{t}^{-1}(x_{t})(\dot{x}_{t}-u_{t,-}(x_{t}))-%
\nabla \cdot u_{t,-}(x_{t})
\end{equation}
with $d_{t}(x)=D_{t}(x,x)$ and $\widehat{u}_{t,+}^{i}=u_{t,+}^{i}-\frac{1}{2}%
\partial _{y^{i}}D_{t}^{ij}(x,y)|_{y=x}$. The time integral in Eq.\thinspace
(\ref{WT}) should be taken in the Stratonovich sense. \vskip0.3cm The
measures $\mu _{0}$ and $\mu _{0}^{\prime }$ in the DFR (\ref{DFR}) do not
have to be normalized or even normalizable. If they are, then distributing
the initial points of the forward and the backward processes with
probabilities $\mu _{0}(dx)$ and $\mu _{0}^{\prime }(x)$, resperctively, we
may define the averages 
\begin{equation}
\left\langle F\right\rangle =\int \mu _{0}(dx)\,{\bm E}_{x}\,F[x],\qquad
\langle F\rangle ^{\prime }\ \equiv \,\int \mu _{0}^{\prime }(dx)\,{\bm E}%
_{x}^{\prime }\,F[x],
\end{equation}
where ${\bm E}_{x}$ (${\bm E}_{x}^{\prime }$) stands for the expectation
value for the forward (backward) process atarting at $x$. \thinspace From
the DFR one may derive a generalisation of the celebrated Jarzynski equality 
\cite{Jarz1,Jarz4}, 
\begin{equation}
\left\langle\,\exp (-W_{\hspace{-0.02cm}_T})\right\rangle =1,  \label{JE}
\end{equation}
which may be viewed as an extension of the fluctuation-dissipation theorem
to the situations arbitrarily far from the equilibrium. Note that the
relation (\ref{JE}) implies the inequality $\langle W_{\hspace{-0.02cm}%
_T}\rangle \geq 0$. \vskip0.3cm To reformulate the DFR in a form where the
entropic interpretation of $W_{\hspace{-0.02cm}_T}$ is clearer, consider the
probability measures $M[dx]$ and $M^{\prime }[dx]$ on the spaces of
trajectories of the forward and and of the backward process, respectively,
such that 
\begin{equation}
\left\langle F\right\rangle \ =\ \int F[x]\,M[dx],\qquad \left\langle
F\right\rangle ^{\prime }\ =\ \int F[x]\,M^{\prime }[dx].
\end{equation}
The DFR may be reformulated in the Crooks form \cite{Crooks2} as the
identity 
\begin{equation}
\left\langle F\exp (-W_{\hspace{-0.02cm}_T})\right\rangle =\langle 
\widetilde{F}\rangle ^{\prime },  \label{rew}
\end{equation}
where $\widetilde{F}[x]=F[\widetilde{x}]$ with $\widetilde{x}_{t}=$ $%
x_{T-t}^{\ast }$, and the relation (\ref{rew}) implies the equality 
\begin{equation}
\widetilde{M}^{\prime }[dx]=\exp (-W_{\hspace{-0.02cm}_T}[x])\,M[dx],
\end{equation}
for the trajectory measures with $\widetilde{M}^{\prime }[dx] =M^{\prime }[d%
\widetilde{x}]$. By introducing the relative entropy $S(M|\widetilde{M}%
^{\prime })=\int \ln (\frac{M[dx]}{\widetilde{M}^{\prime }[dx]})\,M[dx]$ of
the measure $\widetilde{M}^{\prime }$ with respect to $M$, we infer that 
\begin{equation}
\left\langle W_{\hspace{-0.02cm}_T}\right\rangle =S(M|\widetilde{M}^{\prime
}).
\end{equation}
Thus the inequality $\left\langle W_{\hspace{-0.02cm}_T}\right\rangle \geq 0$
follows also from the positivity of relative entropy. One may postulate that 
$\int_{0}^{T}\left\langle J_{t}\right\rangle dt$\ describes the mean entropy
production in the environment modeled by the stochastic noise: 
\begin{equation}
\int_{0}^{T}\hspace{-0.15cm}\left\langle J_{t}\right\rangle dt= \Delta_{%
\hspace{-0.01cm}_T}S_{env}.
\end{equation}
This is coherent with the previous result and particular cases, see \cite
{EYINK,GASP,MAES}. We may then interprete $\int_{0}^{T}\hspace{-0.1cm}
J_{t}dt$ as the fluctuating entropy production in the environment. An easy
calculation leads to the relation 
\begin{equation}
\left\langle W_{\hspace{-0.02cm}_T}\right\rangle =S(\hat{\mu}_{\hspace{%
-0.01cm}_T})-S(\mu _{0})+\Delta_{\hspace{-0.01cm}_T} S_{env}+S(\hat{\mu}_{%
\hspace{-0.01cm}_T}|\mu_{\hspace{-0.01cm}_T}),
\end{equation}
where $\hat{\mu}_{t}(dx)=exp(-\hat{\varphi}_{t}(x))\,dx$ is the measure
describing the time $t$ distribution of the forward process if its initial
distribution were $\mu _{0}(dx)$. $S(\hat{\mu}_{t})=\int \hat{\varphi}_{t}(x)%
\hat{\mu}_{t}(dx)$ is the mean instantenous entropy of the forward process $%
x_{t}$ and $S(\hat{\mu}_{\hspace{-0.01cm}_T})-S(\mu _{0})$ is its change
over time $T $. We could interprete $\hat{\varphi}_{t}(x_{t})$ as the
fluctuating instantenous entropy. In general, $\hat{\mu}_{\hspace{-0.01cm}_T}
$ is not linked to $\mu_{\hspace{-0.01cm}_T}$ of formula (\ref{defu}). The
relative entropy $S(\hat{\mu}_{\hspace{-0.01cm}_T}|\mu_{\hspace{-0.01cm}_T})$
is a penalty due to the use at time $T$ of a measure different than $\hat{\mu%
}_{\hspace{-0.01cm}_T}$. In the case where $\hat{\mu}_{\hspace{-0.01cm}_T}
=\mu_{\hspace{-0.01cm}_T}$, $\left\langle W_{\hspace{-0.02cm}%
_T}\right\rangle $ is the mean entropy production in the system and
environment during time $T$ and we could interpret $W_{\hspace{-0.02cm}_T}$
as the corresponding fluctuating quantity. After a simple calculation \cite
{MNW}, one gets 
\begin{eqnarray}
&&\hspace*{-1cm}\Delta_{\hspace{-0.01cm}_T}S_{env}=\int_{0}^{T} \hspace{%
-0.25cm}\left\langle J_{t}\right\rangle dt=\int_{0}^{T}\hspace{-0.25cm}%
dt\int \big[2\widehat{u}_{t,+}(x)\cdot d_{t}^{-1}(x)\big(\hat{\jmath}%
_{t}(x)dx-u_{t,-}(x)\hat{\mu}_{t}(dx)\big)-(\nabla \cdot u_{t,-})(x)\hat{\mu}%
_{t}(dx)\big],\   \label{moyent} \\
&&\hspace*{-1cm}S(\hat{\mu}_{\hspace{-0.01cm}_T})-S(\mu _{0}) =\int_{0}^{T}%
\hspace{-0.1cm}dt\int \hat{\jmath}_{t}(x)\cdot \nabla \hat{\varphi}%
_{t}(x)\,dx\,,
\end{eqnarray}
where $\hat{\jmath}_{t}$ is the probability current at time $t$ with the
components 
\begin{equation}
\hat{\jmath}_{t}^{i}=\big(\widehat{u}_{t}^{i}-\frac{1}{2}d_{t}^{ij}\partial
_{j}\big)\exp (-\hat{\varphi}_{t})  \label{curr0}
\end{equation}
that satisfies the continuity equation 
\begin{equation*}
\partial _{t}\exp (-\hat{\varphi}_{t})+\partial _{i}\hat{\jmath}%
_{t}^{i}\,=\,0.
\end{equation*}
We shall apply now these results to three type of semiclassical single-mode
laser.

\section{Tuned laser with additive noise}

\subsection{Stationnary case}

Let us consider the most common model of a stationnary laser with no
detuning and with an additive form of the noise \cite{Sargent,Haken}. Its
dynamics is described by the SDE 
\begin{equation}
\frac{dE}{dt}=(a-bE\bar{E})E+\eta ,  \label{stan}
\end{equation}
with $a$ and $b$ real, $b>0$, and with white noise $\eta $ with mean zero
and covariance 
\begin{eqnarray}
\left\langle \eta _{t}\bar{\eta }_{t^{\prime }}\right\rangle &=&D\,\delta
(t-t^{\prime }),  \label{cov} \\
\left\langle \eta _{t}\eta _{t^{\prime }}\right\rangle &=&\left\langle \bar{%
\eta }_{t}\bar{\eta }_{t^{\prime }}\right\rangle \ =\ 0.  \notag
\end{eqnarray}
We can write the covariance matrix in the $(E,\bar{E})$ space as 
\begin{equation}
d=D\big( 
\begin{matrix}
0 & \hspace{-0.1cm}1\cr1 & \hspace{-0.1cm}0
\end{matrix}
\big).
\end{equation}
The equation (\ref{stan}) has then the form of the Langevin equation
describing equilibrium dynamics of the process ${\bm E}_{t}=(E_{t},\bar{E}%
_{t})$: 
\begin{equation}
\frac{d{\bm E}}{dt}=-\frac{1}{2}d\,{\bm\nabla }H_{ab}+{\bm\eta }
\end{equation}
for $H_{ab}(\bm E)=\frac{1}{D}[b(E\bar{E})^{2}-2aE\bar{E}]$. The Einstein
relation is satisfied for the inverse temperature equal to 1 implying that
the Gibbs measure 
\begin{equation}
\mu _{ab}(d\bm E)=Z_{ab}^{-1}\exp (-H_{ab}(\bm E)\,d\bm E  \label{Gibbs}
\end{equation}
is invariant, has a vanishing probability current $\bm j$, and satisfies the
detailed balance 
\begin{equation}
\mu _{ab}(d\bm E_{0})\ P_{0,T}(\bm E_{0};d\bm E)\,=\,\mu _{ab}(d\bm E)\
P_{0,T}(\bm E;d\bm E_{0}).  \label{db1}
\end{equation}
This relation is a particular case of the detailed fluctuation relation (\ref
{DFR}) where the time inversion acts trivially in the spatial sector, i.e. $%
\bm E^{\ast }=\bm E$, the pseudo-vector part of the drift is taken zero, and
we start with the Gibbs measure $\mu _{ab}$ for the forward and the backward
processes. In this case both processes have the same distribution and $W_{%
\hspace{-0.02cm}_T}\equiv 0$. The relation (\ref{db1}) may be projected to
the one for the process $I_{t}=E_{t}\bar{E}_{t}$ describing the the
intensity of the laser: 
\begin{equation}
\mu _{ab}(dI_{0})\ P_{0,T}(I_{0};dI)\,=\,\mu _{ab}(dI)\ P_{0,T}(I;dI_{0}).
\end{equation}
The fluctuating entropy production in the environement may be identified
with the heat production $\Delta_{\hspace{-0.01cm}_T}Q$ which is a state
function here: 
\begin{equation}
\Delta_{\hspace{-0.01cm}_T}Q\, =\,\int_{0}^{T}\hspace{-0.15cm}J_{t}dt\,=\,-H(%
\bm E_{\hspace{-0.01cm}_T}) +H(\bm E_{0}).
\end{equation}
This relations is the first principle of the thermodynamics in the case with
no work applied to the system. If we start with the Gibbs density then the
mean entropy production in the environment $\Delta_{\hspace{-0.01cm}_T}
S_{env}=\left\langle \Delta_{\hspace{-0.01cm}_T}Q\right\rangle $ vanishes (%
\ref{moyent}) as well as the instantaneous entropy production and $W_{%
\hspace{-0.02cm}_T}$. If the process starts with an arbitrary measure $\mu
_{0}(d\bm E)$ then at subsequent times the measure is 
\begin{equation}
\hat{\mu}_{t}(d\bm E)\,=\int \mu _{0}(d\bm E_{0})\ P_{0,t}(\bm E_{0};d\bm E)
\end{equation}
converging at long times to the invariant measure $\mu _{ab}(d\bm E)$.
During this process the mean rate of heat production $\left\langle
q_{t}\right\rangle $ in the environment is (\ref{moyent}) 
\begin{equation}
\left\langle q_{t}\right\rangle =\left\langle J_{t}\right\rangle =-\int ({\bm%
\nabla }H_{ab}\cdot \hat{\bm j}_{t})(\bm E)\,d\bm E\,.
\end{equation}
After an integration by part, this may be written as 
\begin{equation}
\left\langle q_{t}\right\rangle =-\int H_{ab}(\bm E)\ \partial _{t}\hat{\mu}%
_{t}(d\bm E)\,.
\end{equation}

\subsection{Non-stationnary case}

\subsubsection{ Non-stationary net gain coefficient}

Let us consider now the SDE 
\begin{equation}
\frac{dE}{dt}\,=\,(a_{t}-bE\bar{E})E+\eta
\end{equation}
with an explicit time dependence for the (real) net gain coefficient $a_{t}$%
, with $b>0$, and with the white noise $\eta $ as before. The explicit time
dependance $a_{t}$ may result from an external manipulation. In the matrix
notation, the last equation takes the form 
\begin{equation}
\frac{d\bm E}{dt}\,=\,-\frac{1}{2}d\,\bm\nabla H_{t}+\bm\eta \,.
\end{equation}
with $H_{t}\equiv H_{a_{t}b}$. Here, we are outside the scope of the
detailed balance and we enter in the world of transient fluctuation
relations. To find an interesting DFR in this case, let us search for an
appropriate time inversion. For example, we may impose that the backward
process is still described by a Langevin equation but with the hamiltonian $%
H_{t}^{\prime }(\bm E)=H_{t^{\ast }}(\bm E^{\ast })$. By assuming a linear
relation $\bm E^{\ast }=M\bm E$ and by transforming the drift with the
vector rule, we obtain for the drift of the backward process the relation 
\begin{equation}
u_{t}^{\prime }(x)=-\frac{1}{2}MdM^{T}\,(\bm\nabla H_{t}^{\prime })(\bm E).
\end{equation}
To assure that $MdM^{T}=d$, we shall take $M=1$ or $M=D^{-1}d$, i.e. $\bm %
E^{\ast }=\bm E$ or $\bm E^{\ast }=\bar{\bm E}=(\bar E,E)$. In these two
cases, $H_{t}(\bm E^{\ast })=H_{t}(\bm E)$ so that $H_{t}^{\prime }(\bm %
E)=H_{t^{\ast }}(\bm E)$ and the backward process satisfies the same SDE as
the forward process but with the time-dependence of the Hamiltonian
reparametrized. With this choices, a small calculation gives 
\begin{equation}
\int\limits_{0}^{T}J_{t}\,dt\,=\,-\int_{0}^{T}\bm\nabla H_{t}(\bm %
E_{t})\cdot d\bm E_{t}\,=\,-H_{T}(\bm E_{T})+H_{0}(\bm E_{0})+\int_{0}^{T}(%
\partial _{t}H_{t})(\bm E_{t})\,dt\,.
\end{equation}
The first principle of thermodynamics implies then that $\int_{0}^{T}(%
\partial _{t}H_{t})(\bm E_{t})\,dt\,$\ is the work performed on the laser
during a time $T.$ Starting from the Gibbs measure for the forward and the
backward process, we obtain the relation 
\begin{equation}
W_{\hspace{-0.02cm}T}\,=\,-\Delta_{\hspace{-0.01cm}_T}F\,
+\int_{0}^{T}(\partial _{t}H_{t})(\bm E_{t})\,dt\,=-\Delta_{\hspace{-0.01cm}%
_T}F\, -\,\frac{2}{D}\int_{0}^{T}(\partial _{t}a_{t})\,I_{t}\,dt,
\label{Wspe}
\end{equation}
where $\Delta_{\hspace{-0.01cm}_T}F =F_{\hspace{-0.01cm}_T}-F_{\hspace{%
-0.01cm}_0}$ is the change of the Helmholz free energy $\,F_{t}=-\ln \int
\exp (-H_{t}(\bm E))\,d\bm E$. \thinspace The DFR (\ref{DFR}) takes here the
form 
\begin{equation}
\mu _{0}(d\bm E_{0})\ P_{0,T}(\bm E_{0};d\bm E,dW)\,\exp (-W)\,=\, \mu_{%
\hspace{-0.01cm}_T}(d\bm E)\ P_{0,T}^{{\prime }}(\bm E^{\ast };d\bm %
E_{0}^{\ast },d(-W)),  \label{DFRs}
\end{equation}
where $\mu _{t}$ denotes the Gibbs measure corresponding to $H_{t}$. In this
case, there is a non vanishing entropy production in the environnement given
by 
\begin{equation}
\Delta_{\hspace{-0.01cm}_T}S_{env}\,=\,\left\langle \Delta_{\hspace{-0.01cm}%
_T}Q\right\rangle \,=\,\int_{0}^{T}\hspace{-0.2cm}\left\langle
J_{t}\right\rangle dt =\int_{0}^{T}\hspace{-0.2cm}dt\int H_{t}(\bm %
E)\,(\partial _{t}\hat{\varphi}_{t})(\bm E)\,\exp (-\hat{\varphi}_{t}(\bm %
E))\,d\bm E,
\end{equation}
where $\hat{\mu}_{t}(d\bm E)=\exp (-\hat{\varphi}(\bm E))\,d\bm E$ is the
distibution of $\bm E_{t}$ if $\bm E_{0}$ is distributed with the Gibbs
measure $\mu _{0}(d\bm E)$. Note that, in general, $\hat{\mu}_{t}\not=\mu
_{t}$. The associated Jarzynski equality (\ref{JE}) takes the form 
\begin{equation}
\Big\langle\exp \Big[-\int\limits_{0}^{T}(\partial _{t}H_{t})(\bm E_{t})\,dt%
\Big]\Big\rangle\,=\,\exp (-\Delta F),  \label{jarr}
\end{equation}
that is, explicitly, 
\begin{equation}
\Big\langle\exp \Big[\frac{2}{D}\int\limits_{0}^{T}(\partial
_{t}a_{t})\,I_{t}\,dt\Big]\Big\rangle\,=\,\exp \Big(\frac{a_{\hspace{-0.01cm}%
_T}^{2}-a_{0}^{2}}{bD}\Big)\,\frac{1+\mathrm{erfc}(\frac{a_{\hspace{-0.01cm}%
_T}} {\sqrt{bD}})}{1+\mathrm{erfc}(\frac{a_{0}}{\sqrt{bD}})}.  \label{erfc}
\end{equation}
In fact, there is an infinity of Jarzynski equalities that correspond to
different splittings of the drift $\bm u_{t}=-\frac{1}{2}d\bm\nabla H_{t}$
into $\bm u_{t,\pm }$ parts. The peculiarity of the Jarzynski equality with
the functionnal $W_{\hspace{-0.01cm}_T}$ of (\ref{Wspe}) is that upon its
expansion to the second order in the small time variation $a_{t}=a+h_{t}$
with $h_{t}\ll a$ one obtains the standart fluctuation dissipation theorem 
\cite{Han,Risk,Zwan,HaySas,Fal,Che1} 
\begin{equation}
\frac{\delta \left\langle I_{t}\right\rangle }{\delta h_{s}}\Big|_{_{h\equiv
0}}\,=\ \frac{2}{D}\,\partial _{s}\left\langle I_{s}I_{t}\right\rangle _{0}
\label{fdt1}
\end{equation}
for $s\leq t$, where $\left\langle\, {\cdots }\,\right\rangle _{0}$ is the
equilibrium average in the stationary state with $h\equiv 0$.

\subsubsection{External coherent field}

Another frequent way to induce a non-stationary behavior of the laser is to
add an external coherent field at the laser frequency, modulated with a
time-dependant amplitude $E_{t}^{ext},$ which is injected into the cavity 
\cite{Han}. The gain and the self saturation of the laser depends now on the
total field $E_{t}+E_{t}^{ext}$, but the losses depend just of $E_{t}$, so
the equation (\ref{stan}) becomes: 
\begin{equation}
\frac{dE}{dt}=\left( a-b\left| E+E_{t}^{ext}\right| ^{2}\right)
(E+E^{ext})-\alpha E_{t}^{ext}+\eta ,
\end{equation}
where $\alpha $ is the part of the dissipation in the net gain coefficient $%
a.$ This equation takes for $E_{t}^{tot}=E_{t}+E_{t}^{ext}$ the form: 
\begin{equation}
\frac{dE^{tot}}{dt}\,=\left( a-b\left| E^{tot}\right| ^{2}\right)
E^{tot}-\alpha E^{ext}+\frac{dE^{ext}}{dt}+\eta \,.
\end{equation}
Upon denoting $\,-\alpha E_{t}^{ext}+\frac{dE_{t}^{ext}}{dt}=f_{t}$,
\thinspace this may be rewritten as 
\begin{equation}
\frac{d\bm E^{tot}}{dt}\,=\,-\frac{1}{2}d\,\bm\nabla H_{t}+\bm\eta \,.
\end{equation}
with 
\begin{equation*}
H_{t}(\bm E^{tot})\equiv H_{ab}(\bm E^{tot})-\frac{2}{D}\left( \bar{f}%
_{t}E^{tot}+f_{t}\bar{E}^{tot}\right) .
\end{equation*}
In the case where $E_{t}^{ext}$ is not infinitesimal, we are outside the
linear response regime, but the Jarzynski relation (\ref{jarr}) is always
true with 
\begin{equation*}
\partial _{t}H_{t}(\bm E^{tot})=-\frac{2}{D}\left( (\partial _{t}\bar{f}%
_{t})E^{tot}+(\partial _{t}f_{t})\bar{E}^{tot}\right) .
\end{equation*}
In the limit of infint\'{e}simal $f_{t}$, this Jarzynski relation gives once
again the fluctuation dissipation theorem \cite{Han} : 
\begin{eqnarray}
\frac{\delta \left\langle A_{t}\right\rangle }{\delta f_{s}}\Big|_{_{h\equiv
0}} &=&\,\frac{2}{D}\,\partial _{s}\left\langle \bar{E}_{s}^{tot}A_{t}\right%
\rangle _{0}, \\
\frac{\delta \left\langle A_{t}\right\rangle }{\delta \bar{f}_{s}}\Big|%
_{_{h\equiv 0}} &=&\,\frac{2}{D}\,\partial _{s}\left\langle
E_{s}^{tot}A_{t}\right\rangle _{0}.
\end{eqnarray}

\section{Detuned laser with additive noise}

\subsection{Stationary case}

For the stationary case with no tuning \cite{Risken74}, 
\begin{equation}
\frac{dE}{dt}=(a-bE\bar{E})E+\eta ,
\end{equation}
with $\,a=a_{1}+ia_{2}$ and $b=b_{1}+ib_{2}$ complex, $\,b_{2}>0$, and with
covariance of the noise $\eta $ given by Eq.\thinspace (\ref{cov}). The
detuning destroys the Langevin form of the equation because the drift cannot
be put any more in the form $\bm u=-\frac{d}{2}\bm\nabla H$ but, instead, 
\begin{equation}
\bm u=-\frac{d}{2}\bm\nabla H_{a_{1}b_{1}}+iD\,\Pi \bm\nabla H_{a_{2}b_{2}},
\end{equation}
with $\Pi =\big(
\begin{matrix}
_{0} & \hspace*{-0.2cm}_{1}\cr^{-1} & \hspace*{-0.2cm}^{0}
\end{matrix}
\big)$. \thinspace It is easy to see that the probability current of the
Gibbs measure $\mu _{a_{1}b_{1}}(d\bm E)$ is 
\begin{equation}
\bm j(\bm E)=i\,Z_{a_{1}b_{1}}^{-1}\Pi \bm\nabla H_{a_{2}b_{2}}\exp
(-H_{a_{1}b_{1}})=Z_{a_{1}b_{1}}^{-1}(-ib_{2}E^{2}\bar{E}+ia_{2}E,\,-ib_{2}E%
\bar{E}^{2}-ia_{2}\bar{E})\,\exp (-H_{a_{1}b_{1}})
\end{equation}
and that it is conserved: $\bm\nabla \cdot \bm j=0$ because $H$ depends only
on the intensity $I$. It follows that the measure $\mu _{a_{1}b_{1}}(d\bm E)$
is preserved by the dynamics. We are in a steady state \cite{Fal}. The
detailed balance breaks down due to the non-vanishing of current $\bm j$. It
is replaced by the modified detailed balance: 
\begin{equation}
\mu _{a_{1}b_{1}}(d\bm E_{0})\ P_{0,T}(\bm E_{0};d\bm E)\,=\,\mu
_{a_{1}b_{1}}(d\bm E)\ P_{0,T}(\bar{\bm E},d\bar{\bm E_{0}}).
\label{detbalg}
\end{equation}
This relation, once again, implies a detailed balance for the process for
intensity : 
\begin{equation}
\mu _{a_{1}b_{1}}(dI_{0})\ P_{0,T}(I_{0};dI)\,=\,\mu _{a_{1}b_{1}}(dI)\
P_{0,T}(I;dI_{0}).
\end{equation}
The relation (\ref{detbalg}) is a particular case of the DFR (\ref{DFR})
where the time inversion acts in the spatial sector as the complex
conjugation $\bm E^{\ast }=\bar{\bm E}$, with the vector and pseudo-vector
parts of the drift equal to 
\begin{equation}
\bm u_{+}=\big((a_{1}-b_{1}E\bar{E})E,\,(a_{1}-b_{1}E\bar{E})\bar{E}\big)%
\,,\qquad \bm u_{-}=\big(i(a_{2}-b_{2}E\bar{E})E,\,-i(a_{2}-b_{2}E\bar{E})%
\bar{E}\big).
\end{equation}
Here again the backward process that we obtain with this choice of time
inversion has the same distribution as the forward one and the heat
production 
\begin{equation}
\Delta_{\hspace{-0.01cm}_T}Q\, =\,\int_{0}^{T}J_{t}\,dt\,=\,-H_{a_{1}b_{1}}(%
\bm E_{T})+H_{a_{1}b_{1}}(\bm E_{0})
\end{equation}
is a state function. If the forward and the backward processes are
distributed initially with the Gibbs density $\exp(-H_{a_{1}b_{1}})$ then,
in average, there is no entropy production in environment 
\begin{equation}
\Delta_{\hspace{-0.01cm}_T}S_{env}\,=\,\left\langle \Delta_{\hspace{-0.01cm}%
_T}Q\right\rangle =\int_{0}^{T}\hspace{-0.25cm}\left\langle
J_{t}\right\rangle dt\, =\,-\int_{0}^{T}\hspace{-0.25cm}dt\int \bm\nabla
H_{a_{1}b_{1}}\cdot\,\bm j(\bm E)\,d\bm E=-\int_{0}^{T} \hspace{-0.25cm}%
dt\int H_{a_{1}b_{1}}\cdot\,\bm\nabla \bm j(\bm E) \,d\bm E\,=\,0
\label{ent}
\end{equation}
and $W_{\hspace{-0.01cm}_T}=0$. We have the usual features of equilibrium.

\subsection{Non-stationnary case}

Introduction of a time dependence of the net gain coefficient to the
previous model leads to the SDE 
\begin{equation}
\frac{dE}{dt}\,=\,(a_{t}-bE\bar{E})E+\eta
\end{equation}
with an explicit time dependence for the net gain coefficient $%
a_{t}=a_{1,t}+ia_{2,t}$ and $b=b_{1}+ib_{2}$ with $\,b_{2}>0$. Here, the
fluctuation relation can be developed exactly as in Sect.3.2 but now (\ref
{DFRs}) becomes for $\bm E^{\ast }=\bar{\bm E}:$ 
\begin{equation}
\mu _{a_{1,0}b_{1}}(d\bm E_{0})\ P_{0,T}(\bm E_{0};d\bm E,dW)\,=\,\mu
_{a_{1,T}b_{1}}(d\bm E)\ P_{0,T}(\bar{\bm E},d\bar{\bm E_{0}},d(-W)),
\end{equation}
where $\,\mu _{a_{1,t}b_{1}}$ denotes the Gibbs measure corresponding to $%
H_{a_{1,t}b_{1}}$ and 
\begin{equation}
W_{\hspace{-0.01cm}_T}\,=\,-\Delta_{\hspace{-0.01cm}_T}
F_{a_{1}b_{1}}\,+\int_{0}^{T}(\partial _{t}H_{a_{1,t}b_{1}})(\bm %
E_{t})\,dt\,=-\Delta_{\hspace{-0.01cm}_T} F_{a_{1}b_{1}}\,-\,\frac{2}{D}%
\int_{0}^{T}(\partial _{t}a_{1,t})\,I_{t}\,dt,
\end{equation}
The corresponding Jarzynski relation takes the form 
\begin{equation}
\Big\langle\exp \Big[\frac{2}{D}\int\limits_{0}^{T}(\partial
_{t}a_{1,t})\,I_{t}\,dt\Big]\Big\rangle\,=\,\exp \Big(\frac{%
a_{1,T}^{2}-a_{1,0}^{2}}{b_{1}D}\Big)\,\frac{1+\mathrm{erfc}(\frac{a_{1,T}}{%
\sqrt{b_1D}})}{1+\mathrm{erfc}(\frac{a_{1,0}}{\sqrt{b_1D}})},  \label{jars}
\end{equation}
compare to (\ref{erfc}). The second order expansion in the small time
variation $a_{t}=a+h_{t}$ with $h_t=h_{1,t}+ih_{2,t}$ gives now the
fluctuation-dissipation relations 
\begin{equation}
\frac{\delta \left\langle I_{t}\right\rangle }{\delta h_{1,s}}\Big|%
_{_{h\equiv 0}}\,=\ \frac{2}{D}\,\partial _{s}\left\langle
I_{s}I_{t}\right\rangle _{0},\qquad \frac{\delta \left\langle
I_{t}\right\rangle }{\delta h_{2,s}}\Big|_{_{h\equiv 0}}\,=\ 0,  \label{tfd2}
\end{equation}
see \cite{Che1} for the details.

\section{Tuned laser with multiplicative noise}

\subsection{Stationnary case}

It is not always clear a priori whether the noise is better represented by a
multiplicative or additive model. In laser theory, when the randomness is
due to pumping, it is more reasonable to use the multiplicative model of
noise \cite{Sargent}. The stationary laser dynamics is then described by the
non-Langevin SDE for the complex amplitude $E_{t}$ : 
\begin{equation}
\frac{dE}{dt}=(a-bE\bar{E})E+\eta _{t}E,
\end{equation}
with $a$ real, $b$ positive and the white noise $\eta _{t}$ as before. In
complex coordinates, the covariance matrix (\ref{covD}) takes now the form 
\begin{equation}
D(\bm E,\bm E^{\prime })\,=\,\Big( 
\begin{array}{cc}
0 & DE\bar{E^{\prime }} \\ 
DE^{\prime }\bar{E} & 0
\end{array}
\Big)
\end{equation}
and, on the diagonal, 
\begin{equation}
d(\bm E)=D(\bm E,\bm E)=DE\bar E\,\Big( 
\begin{array}{cc}
0 & 1 \\ 
1 & 0
\end{array}
\Big).  \label{cov'}
\end{equation}
One can show directly that the density $\exp (-\varphi(I) )$, where 
\begin{equation}
\varphi (I)\,=\,\frac{2b}{D}I+(1-\frac{2a}{D})\ln {I}
\end{equation}
and $I=E\bar E$, is preserved by the dynamics and corresponds to the
vanishing current, leading to the detailed balance 
\begin{equation}
\exp (-\varphi (\bm E_{0}))\,d\bm E_{0}\ P_{0,T}(\bm E_{0};d\bm E)\,=\,\exp
(-\varphi (\bm E))\,d\bm E\ P_{0,T}(\bm E;d\bm E_{0}).  \label{db3}
\end{equation}
It is normalisable if $a>0$. In this case, the normalized measure $\,\mu (d%
\bm E)=Z^{-1}\exp (-\varphi (I))\,d\bm E\,$ is invariant and we are once
again in an equilibrium case. There is no invariant probability measure when 
$a\leq 0$. Note the the intensity $I$ satisfies here a closed SDE 
\begin{equation}  \label{forint}
\frac{dI}{dt}=2(a-bI)I+(\eta _{t}+\bar\eta_t)I
\end{equation}
that should be taken with the Stratonovich convention.

\subsection{Non-stationnary case}

Introduction of a time dependence of the net gain coefficient to the
previous model results in the SDE 
\begin{equation}
\frac{dE}{dt}=(a_{t}-bE\bar{E})E+\eta _{t}E.  \label{multi}
\end{equation}
With $\bm E^{\ast }=\bm E$ or $\bm E^{\ast }=\bar{\bm E}$ and the vector
rule for the time-inversion of the drift, the backward process solves the
same SDE with $a_{t}$ and $\eta _{t}$ replaced by $a_{t^{\ast }}$ and $%
\eta_{t^{\ast }}$. This time reversal corresponds both to the so called
reversed protocol and to the current reversal of the articles \cite
{CHCHJAR,Che}. The DFR (\ref{DFR}) takes now the form 
\begin{equation}
\exp (-\varphi _{0}(I_{0}))\,d\bm E_{0}\ P_{0,T}(\bm E_{0};d\bm E,dW)\,\exp
(-W)\,=\,\exp (-\varphi _{\hspace{-0.01cm}_T}(I)) \,d\bm E\ P_{0,T}^{\prime
}(\bm E^{\ast };d\bm E_{0}^{\ast },d(-W))  \label{DFRmu}
\end{equation}
with 
\begin{equation}
\varphi _{t}(I)=\frac{2b}{D}I+(1-\frac{2a_{t}}{D})\ln I
\end{equation}
and 
\begin{equation}
W_{\hspace{-0.01cm}_T}=\int_{0}^{T}(\partial _{t}\varphi _{t})(\bm %
E_{t})\,dt\,=\,-\frac{2}{D}\int_{0}^{T}(\partial _{t}a_{t})\ln I_{t}\,dt.
\end{equation}
The intensity process $I_t$ satisfies the SDE (\ref{forint}) with the net
gain coefficient $a$ replaced by $a_t$. \,The backward intensity process is
given by the same SDE with $a_{t}$ and $\eta _{t}$ replaced by $a_{t^{\ast }}
$ and $\eta _{t^{\ast }}$, leading to the DFR (\ref{DFR}) 
\begin{equation}
\exp (-\varphi _{0}(I_{0})))\,dI_{0}\ P_{0,T}(I_{0};dI,dW)\,\exp
(-W)\,=\,\exp (-\varphi _{T}(I))\,dI\ P_{0,T}^{\prime }(I;dI_{0},d(-W)).
\end{equation}
Introducing the distribution of $W_{\hspace{-0.01cm}_T}$ in the forward and
the backward process by the relations: 
\begin{eqnarray*}
P_{0,T}(W)\,dW &=&\frac{\int \exp (-\varphi _{0}(I_{0}))\,dI_{0}\
P_{0,T}(I_{0};dI,dW)dI\,}{\int \exp (-\varphi _{0}(I_{0}))\,dI_{0}}\text{ \
\ \ \ \ \ \ \ \ and } \\
P_{0,T}^{\prime }(W)\,dW &=&\frac{\int \exp (-\varphi _{T}(I_{0}))\,dI_{0}\
P_{0,T}^{\prime }(I_{0};dI,dW)\,dI}{\int \exp (-\varphi _{T}(I_{0}))\,dI_{0}}
\end{eqnarray*}
we obtain by integration (\ref{DFRmu}) the Crooks relation \cite{Crooks1}: 
\begin{equation}
P_{0,T}(W)=P_{0,T}^{^{\prime }}(-W)
\exp (W-\Delta_{\hspace{-0.01cm}_T}F)\text{ \ \ \ \ \ with \
\ \ \ }\Delta_{\hspace{-0.01cm}_T} F=F_{T}-F_{0}\text{,}  \label{Cro}
\end{equation}
where $F_{t}=-\ln\int \exp (-\varphi _{t}(I))\,dI$. \,In the case with
positive $a_{0}$ and $a_{T}$, we may derive the associated Jarzynski
equality: 
\begin{equation}
\big\langle\exp (-W_{T})\big\rangle\,=\,\exp (-\Delta_{\hspace{-0.01cm}_T}F),  \label{JE2}
\end{equation}
where 
\begin{equation}
\exp (-\Delta_{\hspace{-0.01cm}_T}F)
\,=\,\frac{\int \exp (-\varphi _{T}(\bm E))\,d\bm E}{\int
\exp (-\varphi _{0}(\bm E))\,d\bm E}
\end{equation}
or, explicitly 
\begin{equation}
\Big\langle\exp \Big[\frac{2}{D}\int_{0}^{T}(\partial _{t}a_{t})\ln I_{t}\,dt%
\Big]\Big\rangle\,=\,\left( \frac{2b}{D}\right) ^{-2(a_{T}-a_{0})/D}\,\frac{%
\Gamma (2a_{T}/D)}{\Gamma (2a_{0}/D)}.
\end{equation}
Expanded to the second order in $h_{t}=a_{t}-a$, the identity (\ref{JE2})
induces the generalised fluctuation dissipation theorem (for a non-Langevin
case): 
\begin{equation}
\frac{\delta \left\langle \ln I_{t}\right\rangle }{\delta h_{s}}\Big|%
_{_{h\equiv 0}}\,=\ \frac{2}{D}\,\partial _{s}\left\langle \ln I_{t}\,\ln
I_{s}\right\rangle _{0}  \label{fdt3}
\end{equation}
for $s<t$. Once again, it is the fluctuation dissipation theorem associated
to the stochastic equation (\ref{multi}), as it was demonstated in \cite
{Che1}. \vskip 0.3cm \vskip 0.3cm We did a numerical verification of the
Crooks relation (\ref{Cro}) for the case $T=1,$ $a_{t}=1+t,$ $b=1$ and $D=1.$
We realized with Patrick Loiseau\footnote{%
Univ\'{e}rsit\'{e} de Lyon, Ecole Normale Sup\'{e}rieure de Lyon.} a Matlab
computation. Below, we draw $P_{0,1}(W)$, $P_{0,1}^{^{\prime }}(W)$ 
as a function of $W$ and $\ln(\frac{P_{0,1}(W)}{P_{0,1}^{^{\prime }}(-W)})$ 
as a function of $W-\Delta_1F$. The simulation was done on $5000$ 
initial conditions between $0$ and $10.$
For each initial condition, we considered $50$ realizations of the noise.
The interval of discretisation in time was $2^{-15}.$ 

\begin{figure}[!h]
\vskip 0.4cm
\begin{center}
\includegraphics[scale=0.4]{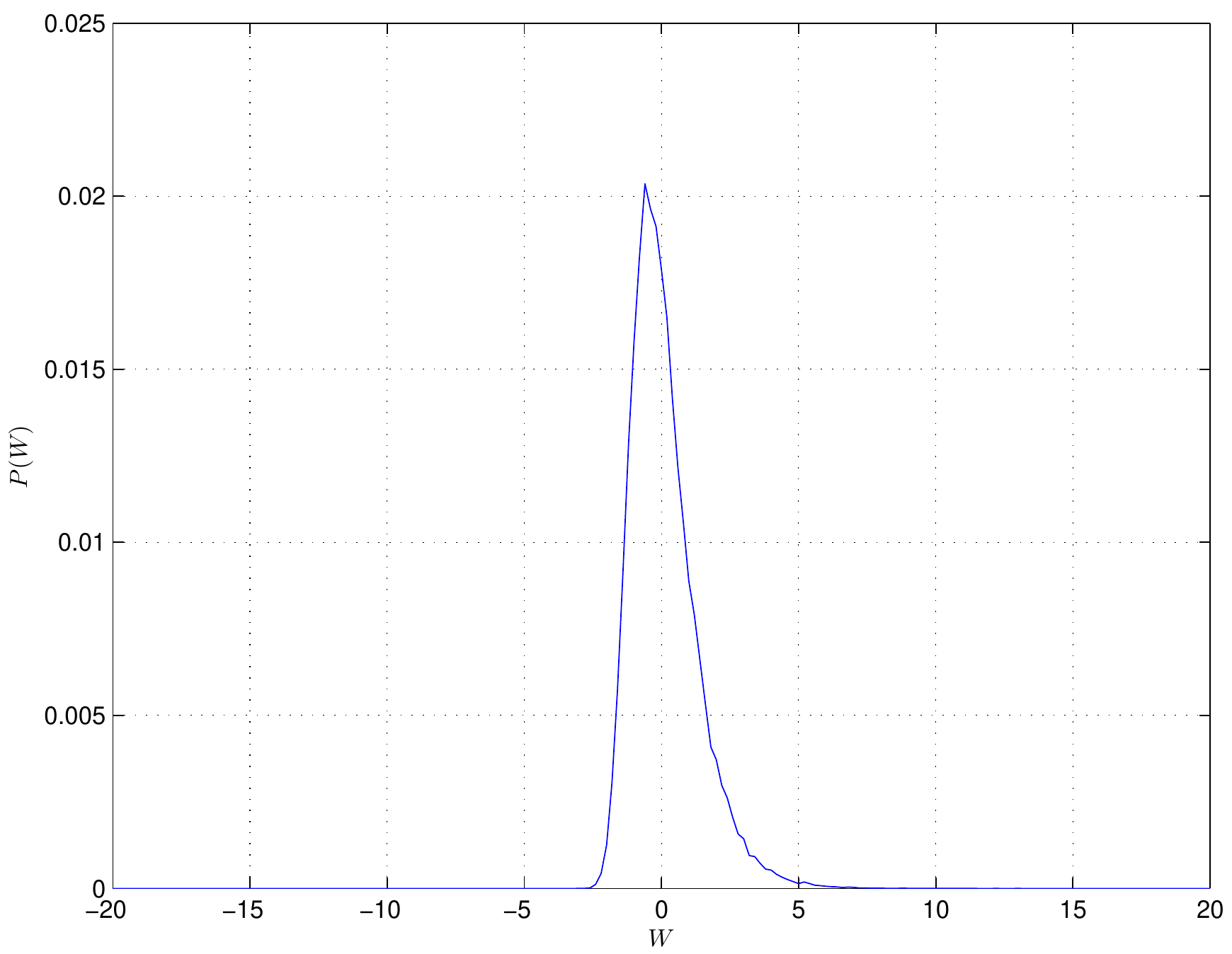}
\end{center}
\caption{$P_{0,1}(W)$ as a function of $W$. Here 
$\langle W_{1}\rangle =0.1023$.}
\end{figure}

\begin{figure}[!h]
\begin{center}
\includegraphics[scale=0.4]{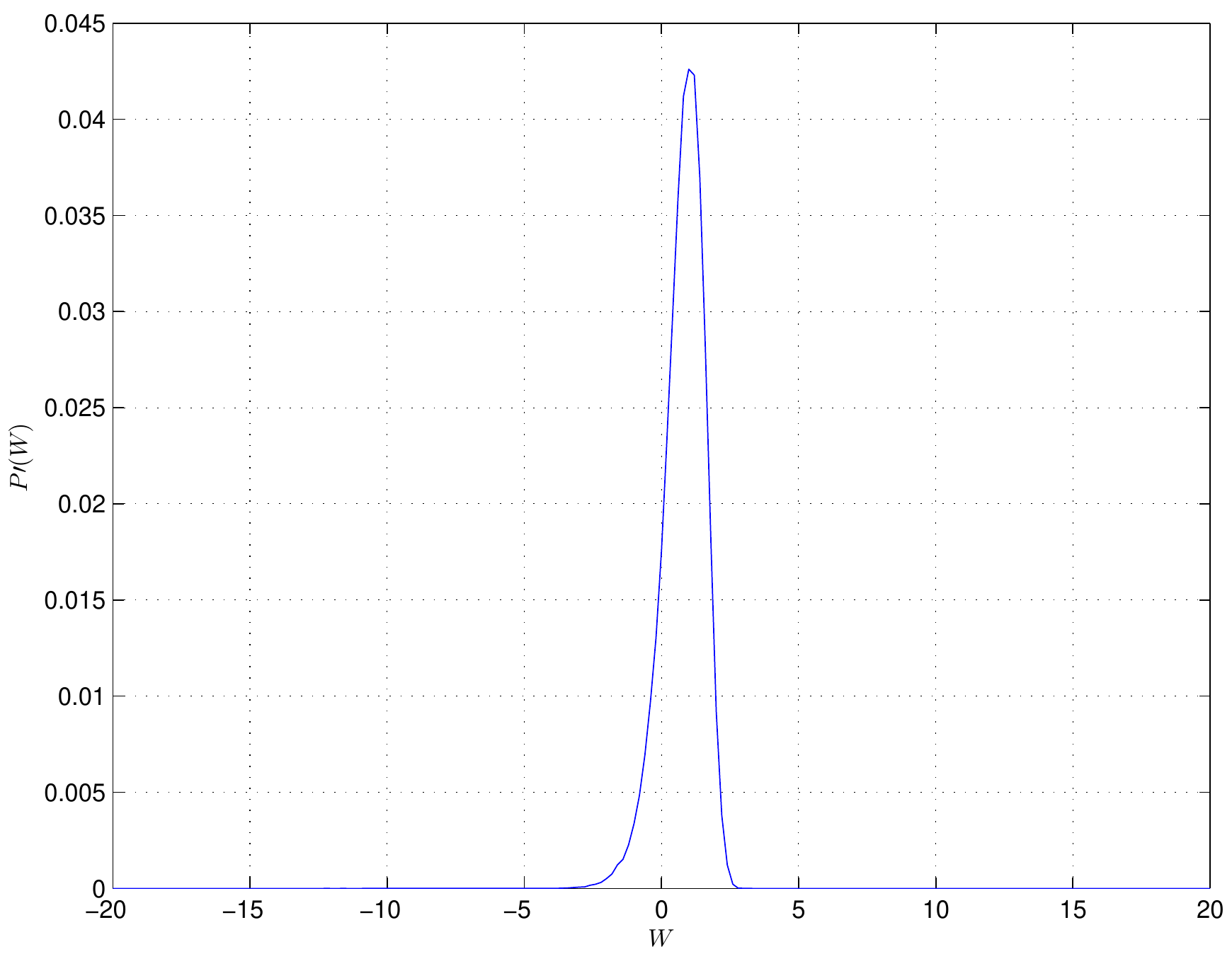}
\end{center}
\caption{$P'_{0,1}(W)$ as a function of $W$. Here 
$\langle W^{\prime}_{1}\rangle = 0.784$.}
\end{figure}

\begin{figure}[!h]
\begin{center}
\includegraphics[scale=0.75]{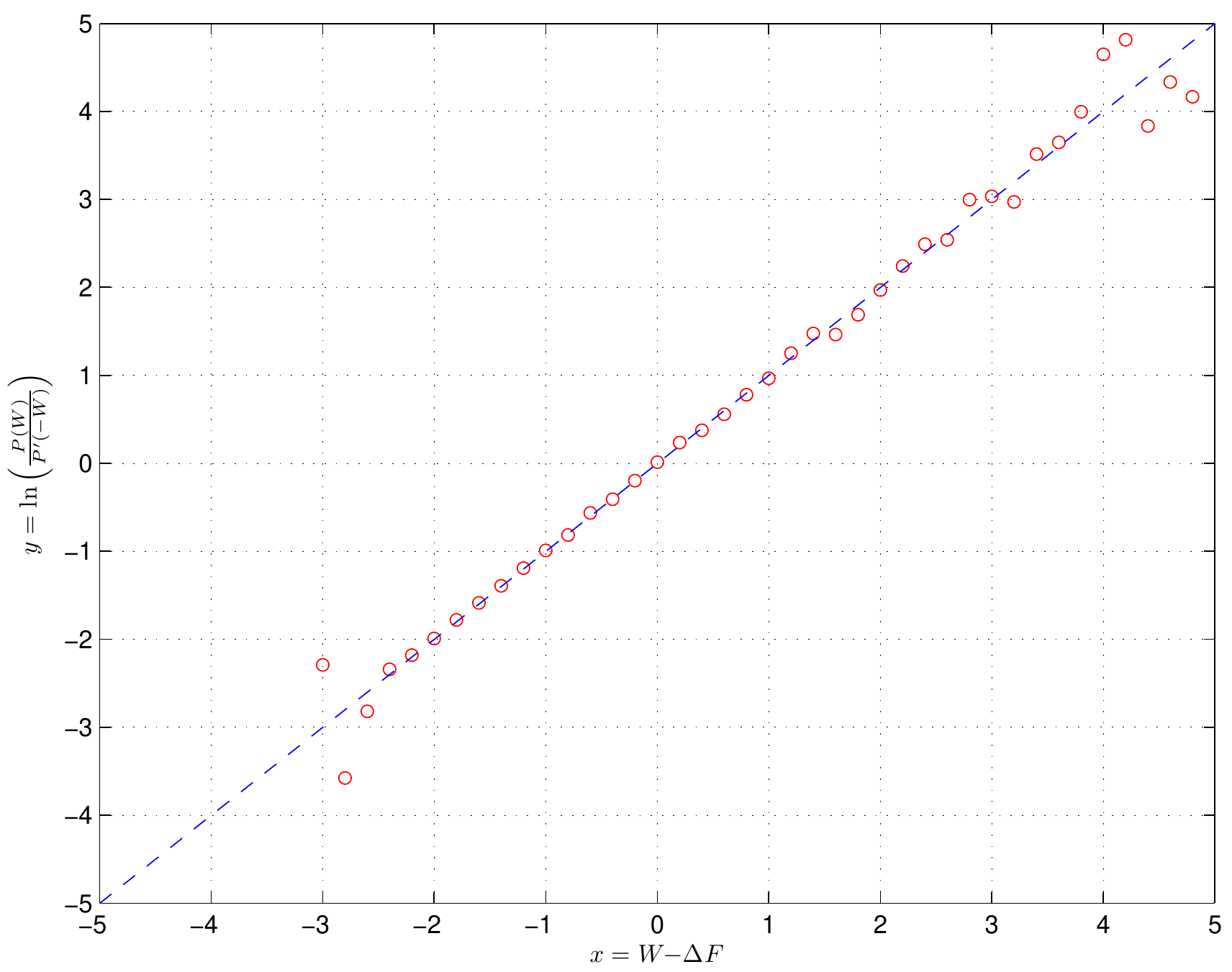}
\end{center}
\caption{$\ln(\frac{P_{0,1}(W)}{P_{0,1}^{^{\prime }}(-W)})$ as a function
of $W-\Delta_1F$. The continuum line is the identity function}
\end{figure}

\newpage

\section{Conclusion}

We have discussed different fluctuation relations for a stochastic model of
the semiclassical regime in a single mode laser. In particular, we showed
that the stationary tuned laser with additive noise has an equilibrium state
with detailed balance (\ref{db1}) and that the detuning preserves the
features (\ref{detbalg}) and (\ref{ent}) of equilibrium. We also studied the
non stationnary case, showing for the tuned and the detuned laser close to
equilibrium the standart fluctuation-dissipation theorems (\ref{fdt1}) and (%
\ref{tfd2}) that extend to the appropriate Jarzynski equality (\ref{jars})
far from equilibrium. Finaly we studied laser with multiplicative noise. We
specified in this case the detailed balance relation (\ref{db3}) satisfied
in the stationary case and the fluctuation-dissipation theorem (\ref{fdt3}).
We also verified numerically the Crooks relation (\ref{Cro}) in the
non-stationary case.


\begin{thebibliography}{99}
\bibitem{CHCHJAR}  Chernyak, V., Chertkov, M., Jarzynski, C.: \textit{%
Path-integral analysis of fluctuation theorems for general Langevin processes%
}. J. Stat. Mech., P08001(2006).

\bibitem{Che}  Chetrite, R., Gawedzki, K. : \textit{Fluctuation relations
for diffusion process}. Commun. Math. Phys., \textbf{282},469-518(2008).

\bibitem{Che1}  Chetrite, R. : Thesis of ENS-Lyon (2008). Manuscript
available at \textit{http://perso.ens-lyon.fr/ raphael.chetrite/}

\bibitem{Crooks1}  Crooks, G. E.: \textit{The entropy production fluctuation
theorem and the nonequilibrium work relation for free energy differences}.
Phys. Rev. \textbf{E 60}, 2721-2726(1999).

\bibitem{Crooks2}  Crooks, G. E.: \textit{Path ensembles averages in systems
driven far from equilibrium}. Phys. Rev. \textbf{E 61}, 2361-2366(2000).


\bibitem{Fal}  Chetrite, R., Falkovich, G., Gawedzki, K. \textit{Fluctuation
relations in simple examples of non-equilibrium steady states.} J. Stat.
Mech. P08005(2008).

\bibitem{EYINK}  Eyink, G. L., Lebowitz, J. L., Spohn, H: \textit{\
Microscopic origin of hydrodynamic behavior: entropy production and the
steady state}. In: Chaos, Soviet-American Perspectives in Nonlinear Science,
ed. D.K. Campbell, pp. 367-397, American Institute of Physics, (1990).

\bibitem{GASP}  Gaspard, P.: \textit{Time-reversed dynamical entropy and
irreversibility in Markovian random processes}. J. Stat. Phys. \textbf{117}
,599-615(2004).

\bibitem{Haken}  Haken, H. : \textit{Laser theory}, Encyclopedia of physics,
Vol. XXV/2c Springer (1970).

\bibitem{Han}  H\"{a}nggi, P. , Thomas, H. : \textit{Stochastic processes :
time evolution, symmetries and linear response}. Physics Reports \textbf{88}%
, 207-319(1982).

\bibitem{HaySas}  Hayashi, K., Sasa, S.: \textit{Linear response theory in
stochastic many-body systems revisited}. Physica \textbf{A 370},
407-429(2006).

\bibitem{Jarz1}  Jarzynski, C.: \textit{A nonequilibrium equality for free
energy differences}. Phys. Rev. Lett. \textbf{78} , 2690-2693(1997).

\bibitem{Jarz4}  Jarzynski, C.: \textit{Hamiltonian derivation of a detailed
fluctuation theorem}. J. Stat. Phys. \textbf{98}, 77-102(2000).


\bibitem{Kurchan}  Kurchan, J.: \textit{Fluctuation theorem for stochastic
dynamics}. J. Phys. \textbf{A 31}, 3719-3729(1998).

\bibitem{LebowSp}  Lebowitz, J., Spohn, H.: \textit{A Gallavotti-Cohen type
symmetry in the large deviation functional for stochastic dynamics}. J.
Stat. Phys. \textbf{95}, 333-365(1999).

\bibitem{MAES}  Maes, C., Nato$\check{\mathrm{c}}$n\'{y}, K.: \textit{Time
reversal and entropy}. J. Stat. Phys. \textbf{110}, 269-310(2003).

\bibitem{MNW}  Maes, C., Nato$\check{\mathrm{c}}$n\'{y}, K., Wynants, B.: 
\textit{Steady state statistics of driven diffusions}. Physica A \textbf{387}%
, 2675-2689(2008).

\bibitem{Mandel}  Mandel, L., Wolf, E. : Optical coherence and quantum
optics. Cambridge Universiry Press, (1995).


\bibitem{Risk}  Risken, H.: The Fokker Planck Equation, second \'{e}dition.
Springer, Berlin-Heidelberg (1989).

\bibitem{Sargent}  Sargent, M., Scully, M.O., Lamb Jr, W.E.: Laser physics.
Reading MA: Addison-Wesley (1974).

\bibitem{Sargent75}  Sargent, M. , Cantrell, C. , Scott, J.F. :\textit{\
Lase-Phase Transition Analogy : Application to first-Order Transitions}.
Optics Comm. \textbf{15}, 13-16(1975).

\bibitem{Risken74}  Seybold, K., Risken, H. : \ \textit{On the theory of a
detuned single mode laser near threshold}. Z. Physics \textbf{267},
323-330(1974).

\bibitem{SpS}  Speck, T., Seifert, U.: \textit{Integral fluctuation theorem
for the housekeeping heat}. J. Phys. A: Math. Gen. \textbf{38},
L581-L588(2005).

\bibitem{Zwan}  Zwanzig, R.: Nonequilibrium Statistical Mechanics. Oxford
University Press (2002).
\end{thebibliography}
\end{document}